\documentclass[aps,prb,superscriptaddress,reprint,twocolumn]{revtex4}

\usepackage{graphicx}
\usepackage{amsfonts,amsmath,amssymb,amsthm}

\usepackage{upgreek}

\newcommand{\MZOZO}{MgZnO/ZnO }
\newcommand{\MZOxZO}{Mg$_x$Zn$_{1-x}$O/ZnO }

\newcommand{\density}{$\times$ 10$^{11}$ cm$^{-2}$}
\newcommand{\mobility}{cm$^2$/Vs }

\newcommand{\Rxx}{$R_{xx}$ }

\usepackage{color}

\begin{document}

\title{Microwave response of interacting oxide two-dimensional electron systems}

\author{D.~Tabrea}
\affiliation{Max-Planck-Institute for Solid State Research, Heisenbergstrasse 1, D-70569
Stuttgart, Germany}

\author{I.~A.~Dmitriev}
\affiliation{Department of Physics, University of Regensburg, 93040 Regensburg, Germany}
\affiliation{Ioffe Physical Technical Institute, 194021 St.~Petersburg, Russia}

\author{S.~I.~Dorozhkin}
\affiliation{Institute of Solid State Physics RAS, 142432 Chernogolovka, Moscow District, Russia.}

\author{B.~P.~Gorshunov}
\affiliation{Moscow Institute of Physics and Technology, Dolgoprudny, Moscow Region 141700, Russia}

\author{A.~V.~Boris}
\affiliation{Max-Planck-Institute for Solid State Research, Heisenbergstrasse 1, D-70569
Stuttgart, Germany}

\author{Y.~Kozuka}
\affiliation{Research Center for Magnetic and
        Spintronic Materials, National Institute for Materials Science, 1-2-1 Sengen, Tsukuba 305-0047,
        Japan}
\affiliation{JST, PRESTO, Kawaguchi, Saitama 332-0012, Japan}

\author{A.~Tsukazaki}
\affiliation{Institute for Materials Research, Tohoku University,
Sendai 980-8577, Japan}

\author{M.~Kawasaki}
\affiliation{Department of Applied Physics and Quantum-Phase
Electronics Center (QPEC), University of Tokyo, Tokyo 113-8656,
Japan} \affiliation{RIKEN Center for Emergent Matter Science
(CEMS), Wako 351-0198, Japan}

\author{K.~von~Klitzing}
\affiliation{Max-Planck-Institute for Solid State Research, Heisenbergstrasse 1, D-70569
Stuttgart, Germany}

\author{J.~Falson}
\affiliation{Max-Planck-Institute for Solid State Research, Heisenbergstrasse 1, D-70569
Stuttgart, Germany}
\email{j.falson@fkf.mpg.de}


\begin{abstract}
 We present an experimental study on microwave illuminated high mobility \MZOZO based two-dimensional electron systems with different electron densities and, hence, varying Coulomb interaction strength. The photoresponse of the low-temperature dc resistance in perpendicular magnetic field is examined in low and high density samples over a broad range of illumination frequencies. In low density samples a response due to cyclotron resonance (CR) absorption dominates, while high-density samples exhibit pronounced microwave-induced resistance oscillations (MIRO). Microwave transmission experiments serve as a complementary means of detecting the CR over the entire range of electron densities and as a reference for the band mass unrenormalized by interactions. Both CR and MIRO-associated features in the resistance permit extraction of the effective mass of electrons but yield two distinct values. The conventional cyclotron mass representing center-of-mass dynamics exhibits no change with density and coincides with the band electron mass of bulk ZnO, while MIRO mass reveals a systematic increase with lowering electron density consistent with renormalization expected in interacting Fermi liquids.
\end{abstract}

\flushbottom
\maketitle

\section{Introduction}

Two-dimensional electron systems (2DES) have been the subject of intense study as they host a remarkably rich set of ground states depending on the strength of the inter-particle interaction. As the charge carrier density $n$ is reduced,  the Coulomb energy ($E_{\mathrm{C}} \propto \sqrt{n}$) becomes comparable and eventually even exceeds the electronic Fermi energy ($E_{\mathrm{F}} \propto n$). In the limit of high concentration, charge carriers interact weakly and the system's parameters follow from band theory. At intermediate densities, a Fermi liquid described by parameters that undergo a renormalization due to interactions, such as the effective mass $m^*$ and the $g$-factor, forms.\cite{COLERIDGE1996, Anissimova2006} Finally, in the dilute limit, a breakdown of the Fermi-liquid paradigm is anticipated. This culminates either in particle localization or, if disorder is sufficiently suppressed, in highly correlated states such as a Wigner crystal.\cite{Wigner1934,Tanatar1989,Knighton2018}

The band effective mass of a 2DES is commonly measured using cyclotron resonance (CR), since in view of Kohn's theorem\cite{Kohn1961} the resonance frequency is insensitive to inter-particle correlations at the vanishingly small momentum of the incident radiation. Estimates of the renormalized effective mass mostly rely on temperature dependent studies of the Shubnikov-de Haas oscillations.\cite{COLERIDGE1996} Recently, oscillatory magnetotransport features that appear under incident microwave radiation, referred to as microwave induced resistance oscillations or MIRO,\cite{ZudovDu2001} have been advanced as an alternative tool for obtaining the interaction-dependent effective mass.\cite{Hatke2013,Shchepetilnikov2017,Fu2017,Shchepetilnikov2018} This method has mainly been deployed in the weakly-interacting regime, where $r_{\mathrm{s}} = E_{\mathrm{C}}/E_{\mathrm{F}} < 2$. \cite{Hatke2013,Shchepetilnikov2017,Fu2017} Recently a different region of parameter space where Coulomb interactions prevail and $r_{\mathrm{s}}$ spans values from 3 to 6  has been accessed. \cite{Shchepetilnikov2018} This was accomplished with \MZOxZO heterostructures which simultaneously posses a low level of disorder. \cite{Falson2011,Falson2016,Falson2018a}  Indeed, state-of-the-art \MZOxZO samples display electron mobilities beyond $10^6$ \mobility as well as quantum lifetimes that are comparable to what the best GaAs heterostructures can offer.\cite{Falson2015b,UMANSKY2009} Accordingly, exotic fractional quantum Hall features have been reported in these samples.\cite{Falson2015a,Falson2018b}

 Here we aim to extract the electron mass in these heterostructures by performing simultaneous magneto-transmission and magnetotransport  measurements under microwave illumination.\cite{Studenikin2005} For the entire span of charge densities, the transmission signal displays resonant features at the cyclotron resonance. An analysis of the density dependence of this signal yields an electron effective mass close to the band mass $m_\mathrm{b} \approx 0.3 \, m_0$ of bulk ZnO,\cite{Baer1967, Button1993} where $m_0$ is the free electron mass. In contrast, the resistively detected magnetotransport signal of the devices exhibits qualitatively different responses depending on the charge carrier density. While low carrier density samples $(n < 3.5$ \density) exclusively display a conventional response due to heating of the electron system during resonant microwave absorption at CR,\cite{Neppl1979,MaanGossard1982,Hirakawa2001} in the higher density regime the response is dominated by the less common MIRO.\cite{Dmitriev2012,Kaercher2016,Monch2020} No CR related feature was detected in the magnetoresistance of high density samples. These two signals permit further analysis of the effective mass. While the CR-associated feature reflects a similar band mass $m_\mathrm{b} \approx 0.3 \, m_0$ to that obtained in transmission studies, the value extracted from MIRO exhibits a systematic increase with decreasing carrier concentration reflecting the renormalization of the Fermi-liquid as interactions augment.\cite{Pudalov2002,Shashkin2003,Tan2005,Kozuka2012} Lastly we provide a plausible explanation for the dominance of the CR response in the photoresistance of low-density samples.

\section{Experiment}

\begin{figure}
\centering
\includegraphics[width=85mm]{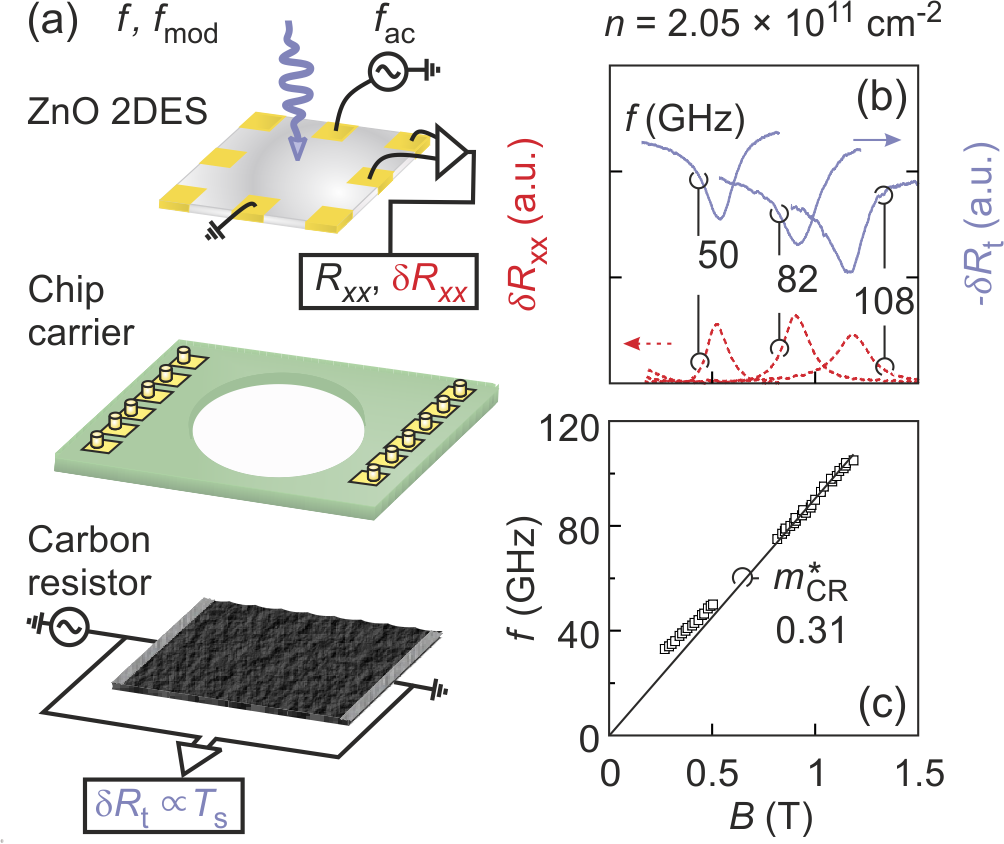}
\caption{
(a) Sketch of the experimental setup. (b) Magnetotransmission data (negated $\delta R_\mathrm{t}(B)$ reflecting the $B$-dependence of the transmittance, $T_\mathrm{s}(B)$, solid lines) and photoresistance $\delta R_{xx}(B)$ (dashed lines) for three microwave frequencies (as marked) obtained at $T = 1.4$ K on a sample with  density $n=2.05$ \density. Curves are shifted vertically for clarity. Linear scales are used. (c) Positions of the minima in the magnetotransmission traces obtained for a number of available microwave frequencies on the sample in panel (b) (open squares). A linear fit crossing the origin for data points corresponding to $f \geq 75$~GHz (solid line) gives the value of the effective mass $m^*_\text{CR}=(0.31 \pm 0.005) m_\mathrm{0}$ associated with the CR in transmission.}
\label{Fig1}
\end{figure}

These studies were performed on a series of \MZOxZO~heterostructures each hosting a 2DES at their heterointerface, with electron densities in the range of $2\leq n\leq 20$ \density\ depending on the Mg content $x$ of the cap layer ($0.01 \leq x \leq 0.15$). Wafers were diced into pieces of approximately $3 \times 3\ {\rm mm}^2$ in order to prepare samples in the van der Pauw geometry with four or eight contacts. The contacts were made by evaporating Ti/Au and/or soldered indium along the perimeter of the sample. The experimental setup is shown in Fig.~\ref{Fig1}(a). Samples are mounted on ceramic chip carriers with a drilled hole of approximately 5 mm in diameter to allow microwave transmission through the sample. A mylar film was glued across this opening to provide support for the sample. Metalized mylar was additionally placed around the perimeter of the chip to limit the transmission of stray radiation. On the backside of the chip carrier, a $4 \times 3$ mm$^2$ carbon-covered kapton film contacted with silver paint was placed. Its resistance $R_\mathrm{t}$ exhibited a strong negative bolometric response $\delta R_\mathrm{t}(B)= - C\, T_\mathrm{s}(B) P_\mathrm{ext}$ proportional to the microwave power $T_\mathrm{s}(B) P_\mathrm{ext}$ transmitted through the sample containing the 2DES. Since both the sensitivity coefficient, $C>0$, and the external microwave power, $P_\mathrm{ext}$,  are $B$-independent, variations of $\delta R_\mathrm{t}(B)$ directly reflect the $B$-dependence of the microwave transmission coefficient $T_\mathrm{s}(B)$. 

The experiments were carried out in a single-shot $^{3}$He cryostat with an axial superconducting coil. The sample is submersed in $^{3}$He liquid and the temperature is varied between 300 mK and 1.4 K by pumping on the $^{3}$He surface. Monochromatic radiation with a frequency of up to 50~GHz was generated using an Agilent 83650 B source. If needed, this signal was additionally amplified and frequency-multiplied to access the $f=75-108$~GHz frequency range. The multiplication leaves an inaccessible window for  $f\approx 50-75$~GHz. The microwaves were delivered to the sample with the help of an oversized rigid rectangular waveguide. Their amplitude was modulated at a frequency $f_\mathrm{mod}=1$~kHz. The longitudinal resistance \Rxx of the 2DEG was measured using low-frequency ($f_{\mathrm{AC}} \approx 10$ Hz) lock-in detection at a bias current of $I=500$~nA. Double modulation using the dual-reference detection capability of an SR860 lock-in amplifier as a fraction of the total signal that is modulated both at $f_\mathrm{AC}$ and at $f_\mathrm{mod}$ was deployed to selectively record weak microwave-induced changes of the low frequency resistance, $\delta R_{xx}$. To improve the signal-to-noise ratio we also relied on double modulation detection of the microwave induced changes of the carbon resistor, $\delta R_{\rm t}$.

\section{Results and Discussion}

\subsection{Transmission measurements}
Solid lines in Fig.\ref{Fig1}(b) display typical magnetotransmission data from the carbon resistor placed below the sample hosting a 2DES with an electron density $n = 2.05$~\density. The change in the carbon resistor value $\delta R_{\rm t}$ has been recorded at different microwave frequencies. A strong maximum was found in each trace, corresponding to a minimum of the transmittance $T_\mathrm{s}(B)$. It is attributed to the CR. Far from the resonance, the sample is nearly transparent to the incoming radiation and the carbon resistor heats up and cools down at the rate of the amplitude modulation of the incident microwave. This translates into a negative and nearly $B$-independent off-resonant signal $\delta R_{\mathrm{t}}$. Near the CR the 2DES absorbs and reflects a larger part of microwaves, which leads to a lower transmitted power and therefore to a lower heating of the carbon resistor. We note that the external radiation power reaching the sample varies significantly with the microwave wavelength. This is due to fluctuations of the incident microwave power caused by the development of standing waves in the waveguide and variations in the output power of the microwave source. Therefore, we employ arbitrary units and refrain from a quantitative comparison of the amplitude for data recorded at different microwave frequencies. We note that the asymmetric line shape of the CR likely originates from interference effects within the sample that depends on the wavelength of the radiation. Over a large frequency range the effect is averaged out.

Panel (c) demonstrates that the $B$-positions of the transmission minima are proportional to the microwave frequency in the high frequency range. The slope obtained from a linear fit of the data for $f\geq 75$~GHz passing through the origin (solid line) establishes that these minima match the CR condition, $f = e B/(2\pi m^*_\mathrm{CR})$, for an effective mass $m^*_\text{CR}=(0.31 \pm 0.005) m_\mathrm{0}$ close to the band mass of ZnO. Due to the finite size of the sample, the 2DES supports a confined plasmon mode. It hybridizes with the cyclotron resonance mode to yield a magnetoplasmon mode of non-zero frequency near $B = 0$. This causes a deviation of the linear $B$-dependence of the observed resonance frequency in the low field limit.\cite{SM} Therefore, data points recorded at frequencies below 50 GHz have been excluded from the mass analysis. The CR is an ubiquitous feature in magnetotransmission for the whole range of electron densities $n = (2 - 20)$~\density~ which we utilize below in gauging the magnitude of mass enhancement obtained from the analysis of photoresistance.

\begin{figure}
\centering
\includegraphics[width=85mm]{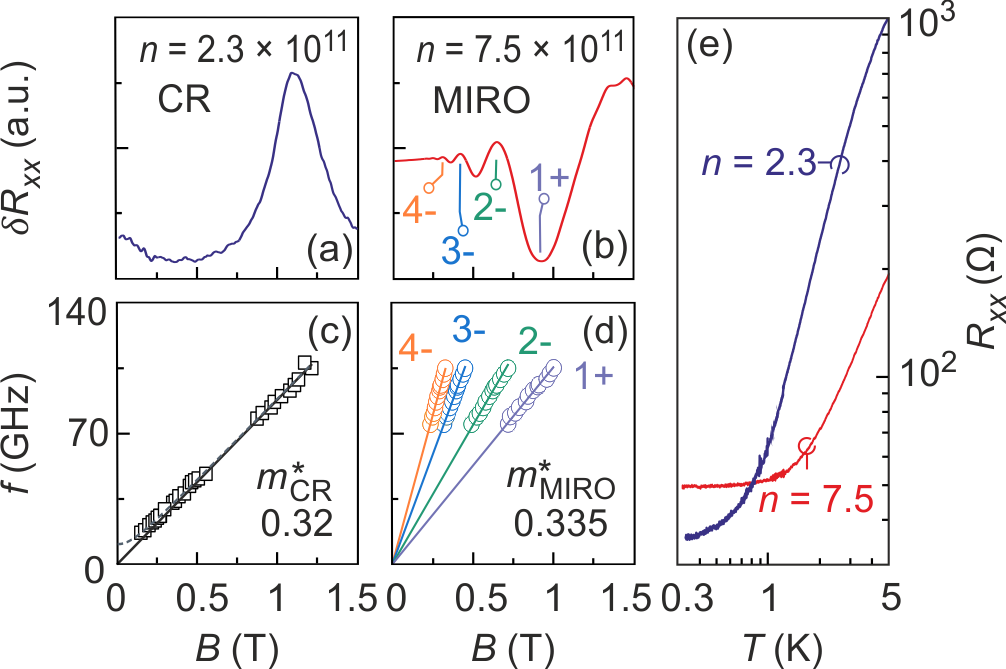}
\caption{Representative examples for the recorded variation  of the longitudinal dc resistance, $\delta R_{xx}$, induced by incident radiation with a frequency $f = 96$ GHz at $T = 1.4$~K. The response differs in samples with low [$n = 2.3$ \density, panel (a)] and high electron density [$n = 7.5$ \density, panel (b)] plotted on a linear scale. (c) Position of the maxima in $\delta R_{xx}$ for the low-density sample shown in panel (a) in the frequency vs. magnetic field plane (open squares) together with a linear fit passing through the origin (solid line). (d) Position of selected MIRO extrema as marked in panel (b) extracted from data obtained at different microwave frequencies. Solid lines are linear fits to the data points using the equation $f = (N \pm 1/4) e B / 2 \pi m^*_{\mathrm{MIRO}}$, with $N = 1, 2, 3$ and $4$. This yields an average value of $m^*_{\mathrm{MIRO}}$ equal to $0.335\ m_0$. (e)  Temperature dependence of the dark resistance $R_{xx}$ at $B=0$ for the structures in panels (a) and (b).
}
\label{Fig2}
\end{figure}

\subsection{Photoresistance measurements}

We now turn our attention to magnetotransport measurements utilizing the double-modulation technique to extract the microwave-induced variation of the longitudinal resistance of the 2DES, $\delta R_{xx}$. In contrast to the magnetotransmission signal that is dominated by the resonant reflection and absorption near the CR for the whole range of electron densities, the resistance measurements manifest more complex behavior depending on the charge carrier density. In low-density samples ($n < 4.3$~\density), the most prominent feature in $\delta R_{xx}$ is a broad peak, as exemplified in Fig.~\ref{Fig2}(a). At $T = 1.4$~K the corresponding change in $R_{xx}$ due to resonant absorption is of the order of a few Ohm. Simultaneous measurement of $\delta R_{xx}$ and of the magnetotransmission signal reveals that the peak is aligned with the minimum in $-\delta R_{\mathrm{t}}$. A comparison between the two sets of data is displayed in Fig.~\ref{Fig1}(b). It is therefore natural to ascribe this peak to a conventional response due to resonant microwave absorption and associated electron heating near the CR. In contrast, high density samples with $n > 4.7$~\density~ display no detectable resonant features in $\delta R_{xx}$ at the CR. Instead, such samples exhibit pronounced $1/B$ periodic magnetooscillations which can be identified as MIRO. A typical trace for higher $n$ is shown in Fig.~\ref{Fig2}(b).

The extracted positions of the maxima in $\delta R_{xx}$ for the low-density sample in Fig.~\ref{Fig2}(a) obtained for different microwave frequencies (open squares) are plotted in Fig.~\ref{Fig2}(c) together with a linear fit passing the origin (solid line). Analogous to the transmission experiment presented above, only the high frequency range $f \geq 75$ GHz was used in the analysis. The slope gives the value of the CR effective mass $m^*_\text{CR}=(0.32 \pm 0.01) m_\mathrm{0}$ which nearly coincides with the value obtained from the minima in the magnetotransmission data of Fig.~\ref{Fig1}(c). This finding reinforces our interpretation of the peak of $\delta R_{xx}$ as an effect of resonant heating of the 2DES in the vicinity of the CR. The extracted value is close to the band mass $m_\mathrm{b} \approx 0.3 \, m_0$ of bulk ZnO.\cite{Baer1967, Button1993} The dashed line in Fig.~\ref{Fig2}(c) illustrates the expected position of the lowest-order magnetoplasmon mode in this sample for a wavelength of the dimensional plasmon equal to twice the sample size, $\lambda_\text{mp}=6$~mm.\cite{SM,Allen1983,Kasahara2012} It demonstrates that finite-size effects are negligible in our large-area samples for frequencies $f$ above 75 GHz. Hence, it is appropriate to describe the observations in terms of the CR in an infinite 2DES.

The period, phase, as well as the damping of the $1/B$-periodic MIRO oscillations observed in high-density samples are all reproduced well by the conventional expression\cite{Dmitriev2012}
\begin{equation}
\delta R_{xx} \propto - \exp{(-\alpha \epsilon)} \sin(2 \pi \epsilon). \label{MIRO deltaR}
\end{equation}
Here $\alpha$ describes the exponential damping at low $B$ [Eq.~(\ref{MIRO deltaR}) is valid for $\alpha \epsilon\gtrsim 1$]. The period of the oscillations is determined by the quasi-particle effective mass $m^*_{\mathrm{MIRO}}$. The latter enters the ratio $\epsilon = \omega/ \omega_{\mathrm{c}}$, where  $\omega = 2\pi f$ is the angular microwave frequency and $\omega_\text{c}=e B/m^*_{\mathrm{MIRO}}$ is the cyclotron frequency determining the distance between neighboring Landau levels for quasi-particles near the Fermi level. The ``bare'' cyclotron mass $m^*_\text{CR}$ extracted from the microwave transmission experiment or the photoresistance feature represents the cyclotron dynamics of the 2DES probed as a whole in the limit $k \rightarrow 0$. \cite{footnotek,KallinHalperin1984} Its value is unaffected by a renormalization of the Fermi liquid in view of momentum conservation and Kohn's theorem.\cite{Kohn1961} In contrast, MIRO involves the scattering of individual quasi-particles at the Fermi surface. The MIRO mass $m^*_{\mathrm{MIRO}}$ is therefore expected to be modified due to renormalization by interactions in a similar way as other transport properties such as  Shubnikov-de Haas oscillations as well as  gap measurements. These probe the electronic system in the opposite limit of large $k$. When a sufficient  number of MIRO harmonics can be resolved in experiment,\cite{SM} the MIRO mass can be determined with high precision by fitting simultaneously the positions of both MIRO minima and maxima to $\epsilon=N\pm1/4$ with integer $N$ (see also Refs.~[\onlinecite{Hatke2013}], [\onlinecite{Shchepetilnikov2017}], and [\onlinecite{Fu2017}]). For the sample in Fig.~\ref{Fig2}(b) the resulting MIRO effective mass is found to be $m^*_{\mathrm{MIRO}}  = (0.335 \pm 0.006) m_0$, i.e.~more than 10\% larger than the cyclotron mass. The open circles in Fig.~\ref{Fig2}(d) shows the positions of several selected extrema of MIRO [as marked in Fig.~\ref{Fig2}(b)] extracted from measurements at different microwave frequencies for illustrative purposes. Solid lines are linear fits using the expression  $f = (N \pm 1/4) e B / 2 \pi  m^*_{\mathrm{MIRO}}$, where $N = 1, 2, 3$ and $4$. An average over the obtained values of the fitting parameter $m^*_{\mathrm{MIRO}}$ yields $0.335\ m_0$.

A plausible reason for the drastically different response to microwave illumination between low [Fig.~\ref{Fig2}(a)] and high density samples [Fig.~\ref{Fig2}(b)] is the much higher temperature sensitivity of the longitudinal resistance in lower density samples. Fig.~\ref{Fig2}(e) displays this temperature dependence for these samples in the absence of radiation and a magnetic field. In both cases the behavior is metallic with a drop in resistance as $T$ is reduced. However, in the higher density sample \Rxx bottoms out for temperatures below approximately $1$~K, whereas in the low-density sample the longitudinal resistance continues to drop down to the lowest accessible temperature. For low density samples, this strong $T$-dependence in the low temperature regime is highly reproducible.\cite{Falson2011} It can be linked to the Bloch-Gr\"uneisen regime for acoustic phonon scattering \cite{Li2013} as well as a higher low-$T$ mobility. The Bloch-Gr\"uneisen regime is entered at a lower temperature in low density samples and alloy or interfacial scattering is weaker due to the reduced Mg-content in the $\rm Mg_xZn_{1-x}O$ cap layer. The response of the 2DES to microwave induced heating can be expressed as $(\delta R_{xx}/ \delta T) \Delta T$,\cite{Neppl1979} and is obviously enhanced when \Rxx shows a higher sensitivity to temperature. The photoresponse is therefore prominent in low density samples, but absent in high density samples.

\begin{figure}
\centering
\includegraphics[width=75.4 mm]{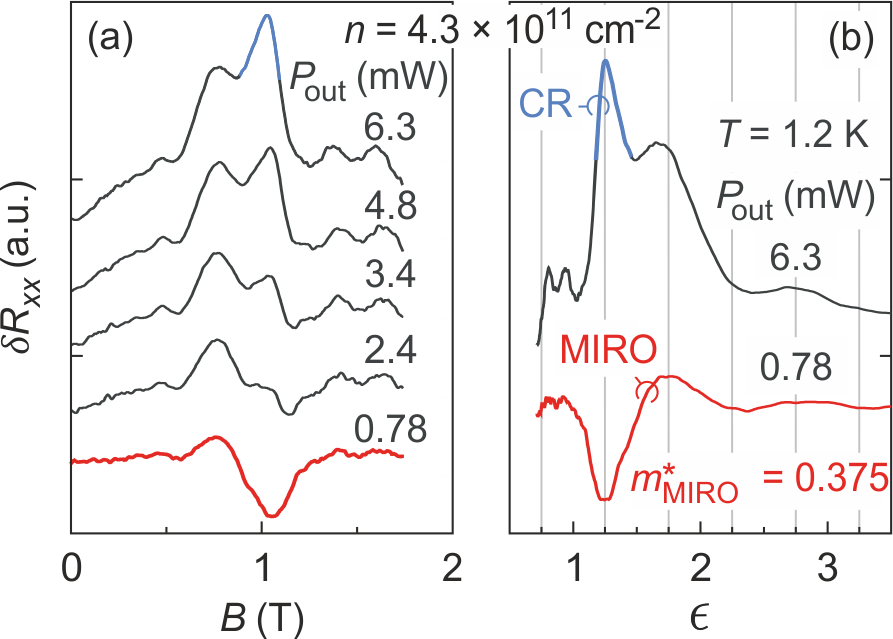}
\caption{Microwave-induced change $\delta R_{xx}$ of the longitudinal resistance recorded on a sample with $n = 4.3$ \density~ for different levels of the output power $P_\text{out}$ (as marked) of the $f  = 95$ GHz microwave radiation. The same data are plotted against $B$ in panel (a) and against $\epsilon= \omega/\omega_\mathrm{c}$ calculated using the MIRO mass $m^*_{\mathrm{MIRO}}  = 0.375$. Linear scales are used.}
\label{Fig3}
\end{figure}

Figure 3 presents $\delta R_{\rm xx}$ data recorded on a sample with an intermediate density $n=4.3$~\density~for different levels of the output power of the microwave generator, $P_\text{out}$, at a fixed frequency of 95 GHz and temperature of the surrounding cryogenic fluid of $T=1.2$~K.
At the highest incident power, $P_\text{out}=6.3$~mW, a strong CR peak appears at the position corresponding to the bare cyclotron mass $m^*_\text{CR}$, as would be expected in a sample that still exhibits a temperature dependence of $R_{xx}$. The CR peak is however accompanied by a MIRO signal. The former decays much faster than MIRO as the microwave power is lowered and heating is suppressed. Indeed, at about an order of magnitude lower power, $P_\text{out}=0.78$~mW, the CR feature has vanished almost entirely, while the MIRO signal remains strong.\cite{SM}
Samples with intermediate densities ($4.3 \leq n \leq 4.7$~\density) therefore enable to simultaneously extract the effective mass unaltered by  interactions as well as the renormalized mass from a single $\delta R_{xx}$ trace. An additional support for this interpretation comes from independently measured transmission signal which provides
the same position of the CR as the CR feature in $\delta R_\mathrm{xx}$. Figure \ref{Fig3}(b) plots $\delta R_{\rm xx}$ as a function of $\epsilon = \omega/\omega_{\mathrm{c}}$ using $m^*_\text{MIRO}=0.375m_\mathrm{0}$ obtained from an analysis of the MIRO at $T = 600$ mK. If the values of the cyclotron and MIRO mass were the same, the CR peak would occur at $\epsilon=1$. However, we see that it coincides with the position of the first MIRO minimum at $\epsilon\simeq 5/4$. We conclude that for this particular density the MIRO mass $m^*_{\mathrm{MIRO}}$ is renormalized by interactions and is approximately 25\% larger than the bare cyclotron mass $m^*_\text{CR}\simeq 0.3\ m_0$.

\begin{figure}
\centering
\includegraphics[width=67.3mm]{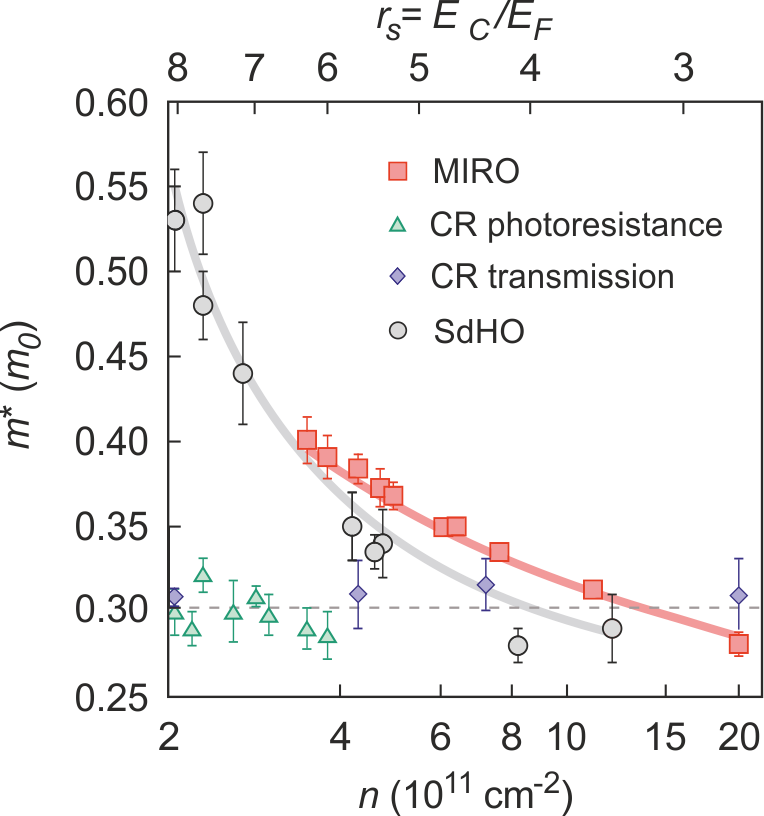}
\caption{The values of effective mass extracted using MIRO period (squares), magnetotransmission (diamonds), CR peak in $\delta R_{xx}$ (triangles), and SdHO (circles) versus the carrier density $n$. Dashed line represents the band mass $m_\mathrm{b} \approx 0.3 \, m_0$ of bulk ZnO, solid lines are guides for the eye.}
\label{Fig4}
\end{figure}

Figure \ref{Fig4} is a compilation of the effective masses obtained via four different methods for samples covering the entire available range of carrier densities. In addition to the CR mass obtained both from magnetotransmission (diamonds) and from the photoresistance $\delta R_{xx}$ (triangles) we include the MIRO mass $m^*_{\mathrm{MIRO}}$ (squares) and the mass $m^*_{\mathrm{SdHO}}$ (circles) obtained from the temperature dependence of the Shubnikov-de Haas oscillations on a set of samples with similar characteristics in previous studies\cite{Falson2015b,Falson2018a}. Within experimental accuracy, the values of the CR mass extracted from the magnetotransmission and from the $\delta R_{xx}$ coincide with each other and with the band effective mass $m_\mathrm{b} \approx 0.3 \, m_0$ (dashed line). The overall mean value for all samples yields  $m^*_{\mathrm{CR}} = (0.3 \pm 0.01)\ m_0$. The MIRO mass  $m^*_{\mathrm{MIRO}}$  was obtained from the dispersion curves $f(B)$ of MIRO extrema, as exemplified in Fig.~\ref{Fig2}(d). Its value displays an increase of 42\% from $0.28 \, m_0$ to $0.4 \ m_0$ as the carrier density $n$ is reduced from 20 \density~ to 3.6 \density. In the range of densities where both methods are applicable, $m^*_{\mathrm{MIRO}}$ agrees fairly well with $m^*_{\mathrm{SdHO}}$.

\section{Conclusions}

In summary, we have presented a combined study of magnetotransport and magnetotransmission on a series of \MZOZO based 2DES under microwave illumination. Across the entire range $2\leq n\leq 20$ \density~ of charge densities the magnetotrasmission displays the CR minima at magnetic field positions consistent with the unrenormalized band mass of the material. The corresponding CR-induced features in magnetotransport were only resolved in low density devices. We identified a strong temperature dependence of the zero-field resistance in such dilute samples, which indicates the reason for a stronger CR response in the photoresistance at low density. MIRO dominate the electrical response in high density samples and reveal a strong renormalization of the quasi-particle effective mass. The reduction at high carrier concentrations as well as the enhancement, which augments as the electron density is diluted, agree with the expected Fermi-liquid renormalization due to interaction effects.

\section*{Acknowledgements}

We thank M. Zudov for useful comments. We acknowledge the financial support of JST CREST Grant Number JPMJCR16F1, Japan. J.F. is grateful for support from the Max Planck-University of British Columbia-University of Tokyo Center for Quantum Materials and the Deutsche Forschungsgemeinschaft (FA 1392/2-1). Y.K. acknowledges JST, PRESTO Grant Number JPMJPR1763, Japan. I.D. acknowledges support from the Deutsche Forschungsgemeinschaft (DM 1/4-1).

\appendix

\section{Determination of the quasiparticle mass from MIRO}

\begin{figure}
\centering
\includegraphics[width=85mm]{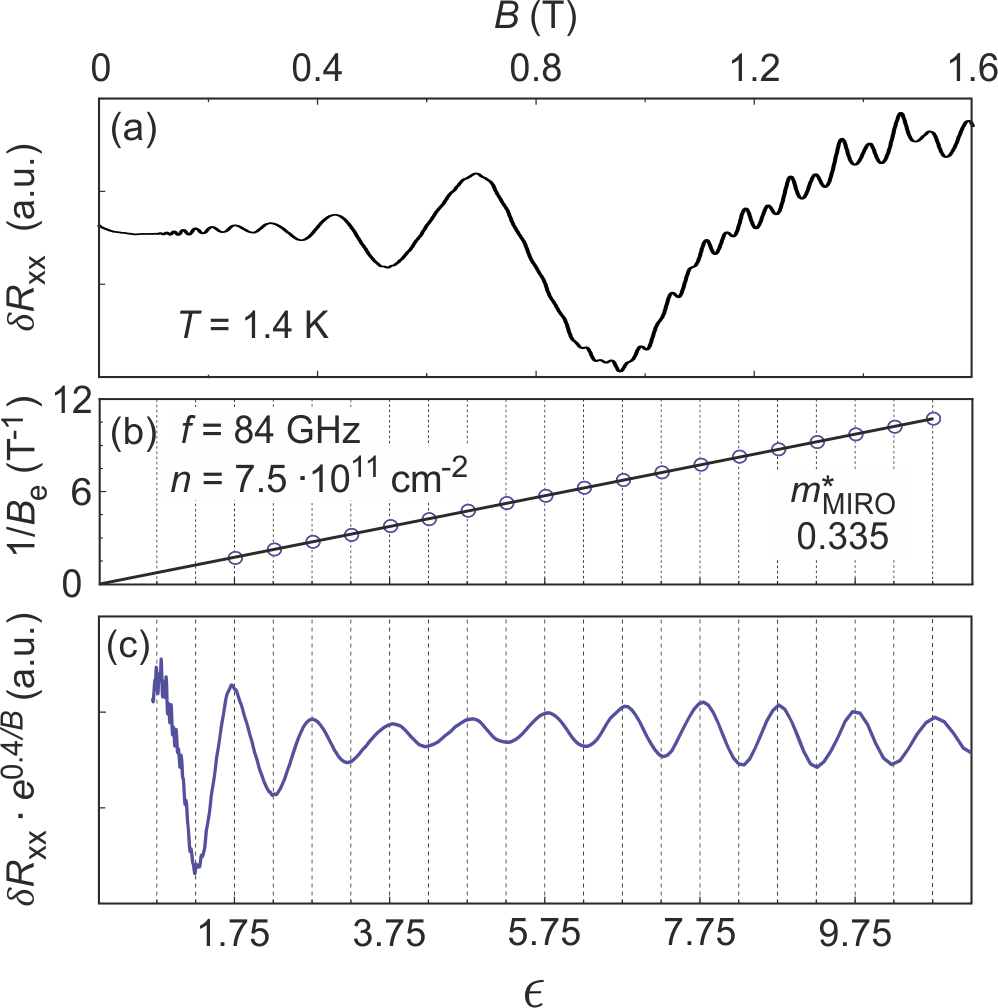}
\caption{Panel (a): Photoresponse $\delta R_{xx}$ for a sample with  $n = 7.5$ \density~ at microwave frequency $f = 84$~GHz. Panel (b): The inverted $B$-positions of MIRO maxima and minima for the data in panel (a) plotted against $N-1/4$ and $N+1/4$, respectively. Here, $N$ is an integer. All points fall on a straight line hitting the coordinate origin. Fitting the slope yields the quasiparticle (MIRO) mass $m^*_\text{MIRO} = (0.335 \pm 0.006) m_0$. In panel (c) the measured microwave-induced change of resistivity $\delta R_{xx}$ (multiplied by $\exp(a/B)$ with $a=0.4$~T for better visibility of high harmonics) is plotted against the inverse of magnetic field which is rescaled to $\epsilon$ using the obtained value of $m^*_\text{MIRO}$. Linear scales are used.}
\label{Fig5}
\end{figure}

The procedure of determination of the effective quasiparticle mass $m^*_\text{MIRO}$ from MIRO is illustrated in Fig.~\ref{Fig5} for a sample with $n = 7.5$ \density. The photoresistance $\delta R_{xx}$ under $f = 84$ GHz microwave illumination is shown in Fig.~\ref{Fig5}(a) as a function of magnetic field. We first extract the positions $B_\mathrm{e}$ of MIRO extrema. In Fig.~\ref{Fig5}(b), the inverse values $1/B_\mathrm{e}$ are plotted against $\epsilon$ assuming a $\mp 1/4$ offset of the MIRO maxima (minima) with respect to the nodes at integer $\epsilon=N$, see Eq.~(1) of the main text. It is seen that within such a representation the data points fall on a straight line going through the axes origin.\cite{footnoteBoffset}
Utilizing the relation $\epsilon = 2 \pi f m^*_{\text{MIRO}}/ e B$, a linear fit with fixed zero intercept yields $m^*_\text{MIRO} = (0.335 \pm 0.006) m_0$. To illustrate the accuracy of the procedure,
in panel (c) we plot full data for $\delta R_{xx}$ against $\epsilon = 2 \pi f m^*_{\text{MIRO}}/ e B$ calculated from the $B$ values using the obtained effective mass. For a better visibility of weak oscillations at high $\epsilon>6$, we multiplied $\delta R_{xx}$ by $\exp(a/B)$ with $a = 0.4$~T. It is seen that all maxima and minima appear precisely at $\epsilon = N \mp 1/4$ for $N > 1$. In this analysis we left out the extrema around $N=1$ where deviations are expected due to a more complex behavior of the MIRO amplitude near the CR.\cite{Dmitriev2012}

As the above example shows, in high-density samples the MIRO effective mass can be accurately determined from a single trace due to the large number of oscillations detectable in the photoresponse. In the low-density regime, however, MIRO are weaker and higher harmonics ($\epsilon > 4$) are not visible. In order to improve the accuracy of extracted $m^*_{\text{MIRO}}$ in this case, we processed data recorded for a larger set of microwave frequencies. In Fig.~\ref{Fig4} we use the average values and standard deviations of $m^*_{\text{MIRO}}$ obtained from the entire collected data set for a given sample.

\section{Power dependence of the microwave response}
\begin{figure}
\centering
\includegraphics[width=85mm]{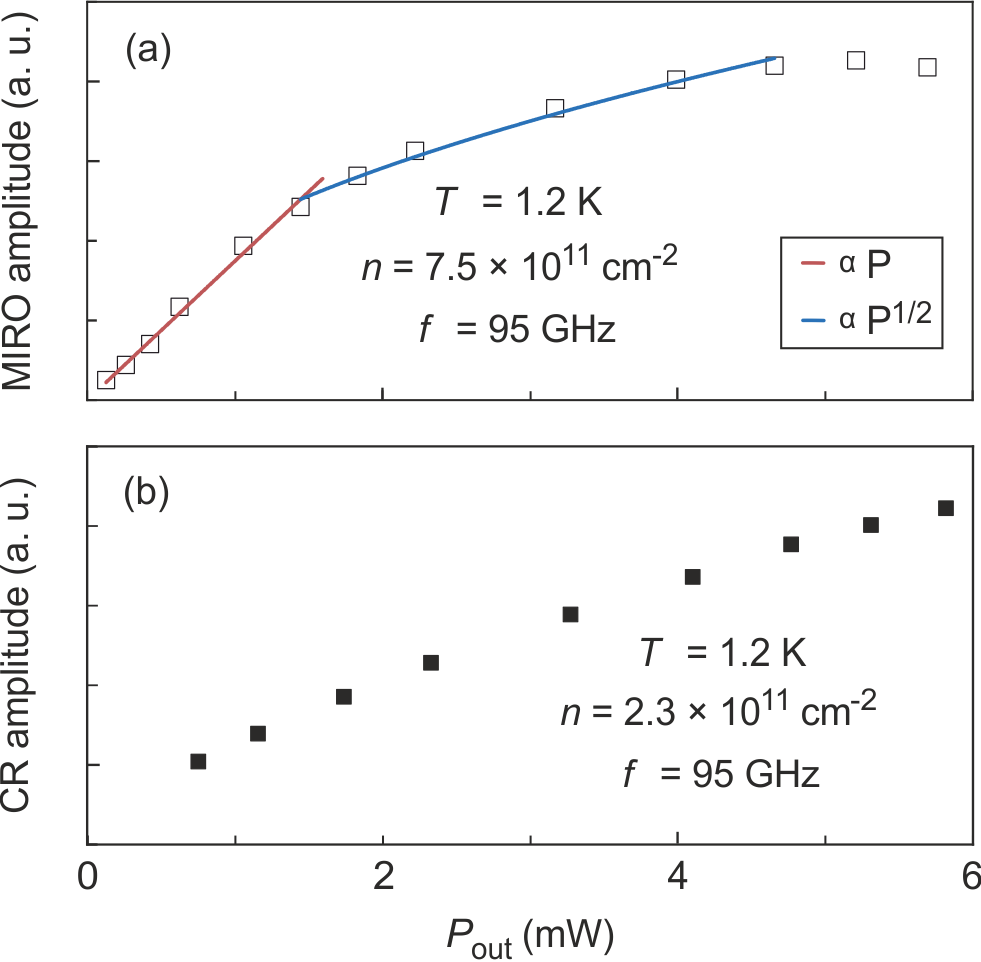}
\caption{Power dependence of the MIRO amplitude [panel (a), sample with $n = 7.5$ \density] and of the amplitude of the CR peak [panel (b), sample with $n = 2.3$ \density]. Solid lines in panel (a) are a guide for the eye illustrating a linear and square-root power dependence. The analysis suggests a transition from a linear to a sublinear regime of MIRO at the output power between 1 and 2 mW. The magnitude of the CR peak in panel (b) shows a monotonic sublinear dependence across the entire power range within which such a signal could be clearly identified.}
\label{Fig6}
\end{figure}

In Fig.~\ref{Fig6} we show the power dependence of the MIRO amplitude [panel (a), sample with $n = 7.5$ \density] and of the amplitude of the CR peak in photoresistance  [panel (b), sample with $n = 2.3$ \density]. Both measurements were made at a temperature $T = 1.2$ K using $f = 95$ GHz radiation. The MIRO amplitude increases linearly in the low-power regime $P < 1.5$  mW. Above this value, a sublinear behavior can be observed. For even higher power radiation ($P > 4 $ mW), the MIRO amplitude saturates and eventually starts to decrease. Importantly, no change of the MIRO phase is observed, i.e. the minima and maxima remain shifted by 1/4 from integer values of $\epsilon$ across the entire available  microwave power range. This suggests that both the transition to the sublinear growth and subsequent decay of MIRO with increasing microwave power are due to heating,\cite{Herrmann2017} and not due to intrinsic nonlinear effects. The latter would rather produce a significant reduction of the MIRO phase and it can even lead to the emergence of additional oscillatory structure around integer $\epsilon$.\cite{Hatke2011,Dmitriev2012,Shi2017} The magnitude of the CR peak in panel (b) shows sublinear growth for power up to the highest available output. At small power, it becomes difficult to isolate the CR peak from the background signal, so, unlike MIRO in panel (a), no clear transition to the linear regime could be identified in this case.

\section{Role of confined magnetoplasmons}

For large microwave frequencies $f>75$~GHz, used for the analysis in the main text, both the magnetic field values where minima in the magnetotransmission [see Fig.~\ref{Fig1} (c)] and maxima in the magnetoresistance response [Fig.~\ref{Fig2} (c)] appear were found to be proportional to the microwave frequency. From the CR condition $f=1/T_c$, with $T_c=2\pi m^*_\mathrm{CR}/e B$, it is possible to extract the CR effective mass $m^*_\mathrm{CR}$. It is found to be close to the band mass of ZnO. At low microwave frequencies $f<50$~GHz we observe a systematic deviation from the linear relationship between the $B$-field at extrema and microwave frequency. Below we show that this deviation is due to a coupling of the cyclotron motion with plasma oscillations in a finite-size 2DES.

Within the simplest model considering a clean 2DES and neglecting electrodynamic retardation effects, the spectum of magnetoplasmons is given by\cite{Ando1982}
\begin{equation}
f_\text{mp}^2 = f_\text{p}^2 + T_\text{c}^{-2}.\label{mp}
\end{equation}
Here the square of the 2DES plasmon frequency,
\begin{equation}
f_\text{p}^2 = \frac{n e^2}{8\pi^2 m^* \bar{\epsilon}}  q,\label{p}
\end{equation}
is proportional to the magnetoplasmon wave vector $q$ and inversely proportional to the effective electron mass $m^*$. In our square-shaped samples with side length $L$, the lowest wave vector corresponding to the fundamental magnetoplasmon mode is $q=\pi/L$. Taking into account that the lateral size $L=3$~mm of the samples significantly exceeds their thickness $w=0.3$~mm (determined by the thickness of the ZnO substrate having  dielectric constant $\epsilon_1=8.5\epsilon_0$ in units of the vacuum permittivity $\epsilon_0$), the effective dielectric constant $\bar{\epsilon}$ entering Eq.~(\ref{p}) can be approximated as \cite{Volkov2010}
\begin{equation}
\bar{\epsilon}=\dfrac{\epsilon_0}{2}+\dfrac{\epsilon_1}{2}\cdot\dfrac{\epsilon_1 \tanh(q w)+\epsilon_0}{\epsilon_1+\epsilon_0 \tanh(q w)}\simeq 2.23 \epsilon_0.\label{ep}
\end{equation}

\begin{figure}
\centering
\includegraphics[width=85mm]{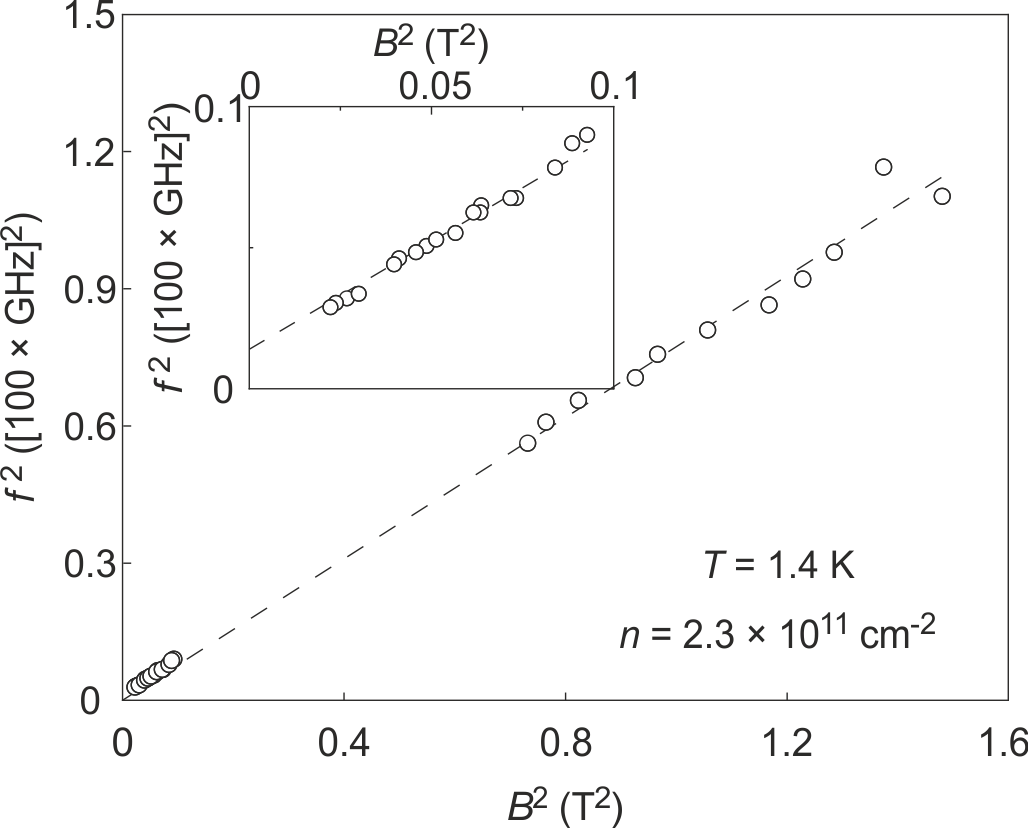}
\caption{The magnetic field values $B$ where a photoresistance peak is detected for different microwave frequencies $f$ on a sample with $n = 2.3$ \density. The data points are plotted using quadratic scales ($f^2$ vs. $B^2$). The main panel presents data for the entire frequency set, while the inset only includes data for $f<30$~GHz. Dashed line in the main plot is a linear fit to the data for $f>75$~GHz with the additional constraint that it passes through the origin. This corresponds to the CR relationship between frequency and field, $f=1/T_c$, with $m^*_\mathrm{CR}=0.32 m_0$. The dashed line in the inset is a linear fit with a non-zero offset, as in Eq.~(\ref{mp}). It yields the plasmon frequency $f_\text{p}=11.8$~GHz.}
\label{Fig7}
\end{figure}

With the help of Eqs. (\ref{mp})-(\ref{ep}) the experimental data can be reproduced well, as illustrated in Fig.~\ref{Fig7} for a sample with density $n = 2.3$ \density. Here we take the magnetic field values $B$ corresponding to the resistance peaks detected for different microwave frequencies $f$, and plot $f^2$ against $B^2$ (open circles). In accordance with Eq.~(\ref{mp}), the data points are found to closely follow a straight line. Within the scale of the main panel (showing data for the entire frequency set) the offset of this line away from the origin [given by $f_\text{p}^2$ according to Eq.~(\ref{mp})] is barely visible. This justifies the application of the simple CR linear relationship between field and microwave frequence,  $f=1/T_c$, for $f>75$~GHz to determine the cyclotron mass in the main text. The corresponding linear fit (obtained for $f>75$~GHz data with $f_\text{p}$ set to zero) is shown as a dashed line in the main panel. The slope corresponds to the cyclotron mass $m^*_\mathrm{CR}=0.32 m_0$ from the main text.

In the inset to Fig.~\ref{Fig7} we present a magnified portion of data for $f<30$~GHz. In the chosen representation, $f^2$ vs. $B^2$, the data points still lie on a straight line, but a positive offset becomes evident. A linear fit of this portion of the data (with slope fixed by $m^*_\mathrm{CR}=0.32 m_0$) yields the offset value $f_\text{p}^2$, corresponding to $f_\text{p}=11.8$~GHz. The same value is obtained from calculation using Eqs.~(\ref{p}) and (\ref{ep}). This good agreement supports the validity of our interpretation of the data in terms of the dimensional magnetoplasmon resonance, although the precise coincidence between calculated and extracted values  of $f_\text{p}$ can also be accidential, in particular, in view of the simplified description of magnetoplasmons used for the above estimates.


\section*{References}
\bibliography{ZnO}

\begin{thebibliography}{43}
\bibitem[{foo({\natexlab{a}})}]{footnotek}
\bibinfo{note}{Here the wave vector $\vec{k}$ characterizes the bound state of
  the excited electron and the resulting hole. In a $B$-field, the wave numer
  $k$ can be written in terms of the intercharge spatial separation $l$ and the
  magnetic length $l_0$ as $k = l/l_0^2$.}
\bibitem[{foo({\natexlab{b}})}]{footnoteBoffset}
\bibinfo{note}{For a more precise fit of the MIRO extrema to $\epsilon = N \mp
  1/4$ for the highest ($N> 6$) MIRO harmonics, it was beneficial to correct
  for a small instrumental shift $B_\text{offset}\sim 1$~mT in the recorded
  values of $B$. For instance, for the data presented in Fig.~\ref{Fig5}, we
  correct all recorded $B$ values by $B_\text{offset}=2$ mT}.
\end{thebibliography}
\clearpage

\end{document}